\documentclass[twocolumn,tradiabstract]{aa} 

\usepackage{natbib}
\usepackage{graphicx}
\usepackage{placeins}
\usepackage{amsmath}
\usepackage{xcolor}
\usepackage{geometry}
\newcommand\mancha{\textsc{Mancha3D-2F~}}

\begin{document} 

\title{Magnetic field amplification and structure formation by the Rayleigh-Taylor instability.}
\author{B. Popescu Braileanu
          \inst{1,2,3}
		\thanks{ \email{\\beatriceannemone.popescubraileanu@kuleuven.be}}
          \and
          V. S. Lukin\inst{4}
          \and
          E. Khomenko\inst{2,3}
}
          
\titlerunning{Rayleigh-Taylor instability}
\authorrunning{Popescu Braileanu et al.}

\institute{
Centre for mathematical Plasma Astrophysics, KU Leuven, 3001, Leuven, Belgium
\and Instituto de Astrof\'{\i}sica de Canarias, 38205 La Laguna, Tenerife, Spain
\and Departamento de Astrof\'{\i}sica, Universidad de La Laguna, 38205, La Laguna, Tenerife, Spain
\and National Science Foundation, Alexandria, VA, 22306, USA}

\date{Received 2021; Accepted XXXX}
 
\abstract{
We report on results of high resolution two fluid non-linear simulations of the magnetized Rayleigh Taylor Instability (RTI) at the interface between a solar prominence and the corona.  These follow results reported earlier by  \cite{rti1,rti2} on linear and early non-linear RTI dynamics in this environment.  { The focus of this paper is on generation and amplification of magnetic structures by RTI.}  The simulations use a two fluid model that includes collisions between neutrals and charges, including ionization/recombination, energy and momentum transfer, and frictional heating. The 2.5D magnetized RTI simulations demonstrate that in a fully developed state of RTI a large fraction of the gravitational energy of a prominence thread can be converted into quasi-turbulent energy of the magnetic field.  RTI magnetic energy generation is further accompanied by magnetic and plasma density structure formation, including dynamic formation, break-up, and merging of current sheets and plasmoid sub-structures.  The flow decoupling between neutrals and charges, as well as ionization/recombination reactions, are shown to have significant impact on the structure formation in magnetized RTI.}


\maketitle
%

\section{Introduction.}

Rayleigh Taylor Instability (RTI) is an instability which occurs in many astrophysical contexts when a heavier fluid is accelerated against a lighter one. In the Sun, RTI is routinely detected in high-resolution observations of solar prominences. This instability manifests itself as small scale upflows and downflows at the interface between prominences and the corona \citep{Berger_2017}. Several numerical models of the magnetic RTI for solar conditions have been reported in the literature \citep[see e.g., ][]{2012ApJ...756..110H, 2012ApJ...746..120H, 2015ApJ...806L..13K, 2016ApJ...820..125T}. These models, done under magnetohydrodynamic (MHD) approximation, allowed to access the non-linear stages of the instability and study the plasma mixing, reconnection and the associated dynamical events.

{ However, solar prominences are composed of partially ionized chromospheric material, which is much denser and cooler than the surrounding corona \citep{2010Labrosse}. The collisional timescale between neutrals and charges in prominences is comparable to the hydrodynamic timescale. Therefore, single-fluid MHD modeling of prominence RTI is not sufficient to describe the physics at small scales and multi-fluid models of RTI are required.} In our recent work \citep{rti1, rti2}, RTI was modeled in the two-fluid approximation in a configuration proposed initially by \citet{2012Leake}. These simulations have demonstrated how the RTI cut-off can be affected by ion-neutral effects, and have shown the complex non-linear dynamical phenomena due to decoupling between the charged and neutral components. 
The aim of the current paper is to study in detail the dynamics and the structure formation associated with non-linear { magnetized} RTI development.



{ The impact of the magnetized RTI dispersion relation for a given system on its non-linear development in a single-fluid approximation has been previously studied by \citet{Hillier2016}. During the non-linear phases of the instability, the in-plane magnetic field lines are stretched and the field is amplified \citep{jun,rti-tub} as the mixing of the fastest growing modes becomes turbulent. The magnetic energy dominates at small scales \citep{Hillier2016}, where the dynamics becomes increasingly important as the nonlinear phase evolves; these are also the scales most impacted by the two-fluid ion-neutral effects.

The turbulent mixing of the magnetized plasma leads to magnetic field lines of opposite polarity being brought together by the flow, leading to creation of current sheets and magnetic reconnection \citep{Cowley1997}.} Because the value of the Ohmic resistivity coefficient in the solar corona is very small, this nonideal effect is only relevant at very small scales, where it is possible to have very high values of current density concentrated in thin current sheets \citep{rec3d}.
Observations have shown bidirectional jets emerging from null points in the corona and photosphere \citep{obsrec2,obsrec3}, and in prominences \citep{obsrec1} with velocities similar to the Alfv\'en speed, consistent with basic reconnection theory. The current sheets can become unstable creating secondary magnetic structures often referred to as plasmoids \citep{Shibata2001}. 

There have been a number of studies where magnetic reconnection was investigated using a two-fluid ion-neutral model under controlled and idealized setups. It has been demonstrated that, when ions and neutrals are allowed to have independent dynamics, they can collisionally decouple in inflows, but are still coupled in outflows \citep{2012Leake}. Two-fluid effects have been shown to increase the reconnection rates compared to a classical Sweet-Parker model \citep{2012Leake,Ni2018a} in some collisionality regimes. The incomplete collisional coupling between charges and neutrals creates thinner current sheets relative to the single-fluid MHD approximation, as only the charged particles are influenced by the magnetic forces on  scales below the neutral-ion collisional mean free path.
The two-fluid simulations show faster thinning of the current sheets and increased growth of the plasmoids compared to the MHD case \citep{Murtas_2021}.
The formation of plasmoids in the current sheet, and to some extent the Hall effect, also enhance the reconnection rates \citep{2013Leake,Ni2018a,Ni2018b}; while ionization-recombination reactions and radiative losses can have significant impact on the thermodynamic properties of a reconnecting current sheet \citep{Ni2018a}. 

We note that the vast majority of the numerical magnetic reconnection studies cited above have focused on exploring the physics of a single pre-formed reconnection region; or a magnetic configuration pre-configured to form a single reconnection region.  Here we explore a different situation where the reconnecting current sheets are naturally created by the dynamics associated with RTI.  { To our knowledge, this is the first exploration of self-consistently emergent turbulent magnetic reconnection in the two-fluid partially ionized regime associated with RTI in solar prominences.

In the following sections} we describe a high resolution simulation of RTI in a 2.5D geometry with parameters corresponding to a solar prominence thread\footnote{A structure which is thin in one dimension becomes  a sheet in a 2.5D geometry, and actually some of the threads observed in prominences might be sheets, projected in the plane of sky.}. The magnetic field of approximate strength of 10~G is directed out of the perturbation plane, inclined by 1 degree and sheared over 1~Mm scale. There is a smooth transition between the prominence thread and the corona.  The two-fluid equations and the detailed initial conditions are presented in Section~\ref{section2_eq}. We describe the overall dynamics of the RTI, and the magnetic field amplification in Section~\ref{section3}. We analyze formation of reconnection current sheets, secondary plasmoid instabilities, and the resulting spectra of magnetic and plasma density structures in Section~\ref{section4}. We present our conclusions in Section~\ref{section5}. 

{
\section{Description of the problem}
\label{section2_eq}
The two-fluid equations solved numerically by the \mancha code are \citep{Popescu+etal2018},
\begin{align}  \label{eq:continuity}
&\frac{\partial \rho_n}{\partial t} + \nabla \cdot (\rho_n\vec{v}_n) = S_n, \nonumber\\ 
&\frac{\partial \rho_c}{\partial t} + \nabla \cdot (\rho_c\vec{v}_c) = -S_n,  \nonumber \\
&\frac{\partial (\rho_n\vec{v_n})}{\partial t} + \nabla \cdot (\rho_n\vec{v_n} \otimes \vec{v_n} +{\bf\hat{p}_n} )
= -\rho_n\vec{g} +\vec{R}_n,  \nonumber \\
&\frac{\partial (\rho_c\vec{v_c})}{\partial t} + \nabla \cdot (\rho_c\vec{v}_c\otimes\vec{v}_c + {\bf\hat{p}_c} )
=\vec{J}\times\vec{B} - \rho_c\vec{g}  -\vec{R}_n\,,\nonumber \\
&\frac{\partial}{\partial t} \bigl( e_n +\frac{1}{2}\rho_n v_n^2 \bigr) + \nabla \cdot \bigl[ \vec{v}_n (e_n + 
\frac{1}{2}\rho_n v_n^2 )  + {\bf\hat{p}_n} \cdot \vec{v}_n \nonumber \\ & - \ K_n \vec{\nabla} T_n\bigr ] = -\rho_n\vec{v}_n \cdot \vec{g}  + M_n,  \nonumber \\
&\frac{\partial}{\partial t} \bigl(  e_c+\frac{1}{2}\rho_c v_c^2 \bigr) +  \nabla \cdot \bigl[ \vec{v}_c (e_c + 
\frac{1}{2}\rho_c v_c^2 )  + {\bf\hat{p}_c} \cdot \vec{v}_c  \bigr] \nonumber  \\&= -\rho_c\vec{v}_c \cdot \vec{g} + \vec{J} \cdot \vec{E} -M_n\,,\nonumber \\
&\frac{\partial \vec{B}}{\partial t} = \nabla \times \bigl(  \vec{v}_c \times \vec{B} \bigr),
\end{align}

\noindent
where we note that we utilize the ideal Ohm's Law for evolving the magnetic field.  The collisional terms $S_n$, $\mathbf{R}_n$ and $M_n$ which appear in the continuity, momentum and energy equations, respectively are,
\begin{eqnarray} \label{eq:s}
S_n = \rho_c \Gamma^{\rm rec} - \rho_n\Gamma^{\rm ion},\nonumber \\
\vec{R}_n = \rho_c \vec{v}_c \Gamma^{\rm rec}  - \rho_n \vec{v}_n \Gamma^{\rm ion} 
+ \alpha \rho_n \rho_c (\vec{v}_c - \vec{v}_n),\nonumber \\
M_n = \left ( \frac{1}{2} \Gamma^{\rm rec} \rho_c v_c^2  - \frac{1}{2}\rho_n v_n^2 \Gamma^{\rm ion} \right )  
\nonumber \\ + \frac{1}{\gamma-1} \frac{k_B}{m_n} \left ( \rho_c T_c \Gamma^{\rm rec} - \rho_n T_n \Gamma^{\rm ion} \right) \nonumber \\ 
+ \frac{1}{2} ({v_c}^2 - {v_n}^2) \alpha \rho_c \rho_n  
\nonumber \\
+\frac{1}{\gamma-1} \frac{k_B}{m_n}(T_c - T_n)\alpha \rho_c \rho_n\,.
\end{eqnarray}
Expressions for $\Gamma^{\rm ion}$ and $\Gamma^{\rm rec} $as functions of $n_e$ and $T_e$ are given in \cite{1997Voronov} and \cite{2003Smirnov}, \citep[see also Appendix in ][]{Popescu+etal2018}:

\begin{equation}
\Gamma^{\rm rec}  \approx \frac{n_e}{\sqrt{T_e^*}}  2.6 \cdot 10^{-19}; \,\,\,\, {\rm s^{-1}}
\end{equation}

\begin{equation}
\Gamma^{\rm ion}  \approx n_e A \frac{1}{X + \phi_{\rm ion}/{T_e^*}}\left(\frac{\phi_{\rm ion}}{T_e^*}\right)^K  e^{-\phi_{\rm ion}/T_e^*}; \,\,\,\,  {\rm s^{-1}}
\end{equation}
where
$\phi_{\rm ion} = 13.6 eV$,
$T_e^*$ is electron temperature in eV,
$A = 2.91 \cdot 10^{-14}$,
$K$ = 0.39, and $X$ = 0.232.

\noindent The  collisional parameter $\alpha$  combines the elastic collisions and the charge-exchange interactions:
\begin{equation}
\alpha = \alpha_{\rm el} + \alpha_{cx}\,.
\end{equation}
The elastic collisions include both collisions between ions and neutrals and between electrons and neutrals:
\begin{equation} \label{eq:alpha_el}
\alpha_{\rm el} = \frac{m_{in}}{{m_n}^2} \sqrt{ \frac{8 k_B T_{cn}}{\pi m_{in}}}  \Sigma_{in} + \frac{m_{en}}{{m_n}^2} \sqrt{ \frac{8 k_B T_{cn}}{\pi m_{en}}}  \Sigma_{en}\,.
\end{equation}

\noindent
In the above Eq.~(\ref{eq:alpha_el}), $T_{\alpha\beta}= (T_\alpha +T_\beta)/2$ is the average temperature,  and $m_{\alpha\beta } = m_\alpha m_\beta/(m_\alpha + m_\beta ) $ is the reduced mass of particles $\alpha$ and $\beta$.

\noindent The charge-exchange (elastic) collisional parameter
between particles of species $\alpha$ (ions) and particles
of species $\beta$ (neutrals), is approximately expressed as \citep[see, e.g.][]{2012Meier}:
\begin{equation}  \label{eq:alpha_cx}
\alpha_{cx} = \frac{1}{m_H}\left (V^{cx}_0 + V^{cx}_{1\alpha} + V^{cx}_{1\beta} \right) \Sigma_{cx}
\end{equation}
\noindent with:
\begin{eqnarray}
V^{cx}_0 &=& \sqrt{\frac{4}{\pi} {{v_T}_\alpha}^2 + \frac{4}{\pi} {{v_T}_\beta}^2   + {{v_{\rm rel}}_{\alpha\beta}}^2} \nonumber \\
V^{cx}_{1\alpha} &=& {{v_T}_\alpha}^2  \left( 4 (\frac{4}{\pi} {{v_T}_\beta}^2 + {{v_{\rm rel}}_{\alpha\beta}}^2 ) +
\frac{9 \pi}{4} {{v_T}_\alpha}^2    \right)^{-0.5}
\nonumber \\
\Sigma_{cx} &=& 1.12 \times 10^{-18} - 7.15 \times 10^{-20} \text{ln} (V^{cx}_0)
\end{eqnarray}
\noindent where ${v_T}_\alpha$, and ${v_{\rm rel}}_{\alpha\beta}$ are the thermal velocity of particles of species $\alpha$,
and the module of the relative velocity between particles of species $\alpha$ and particles of species $\beta$, respectively:
\begin{eqnarray}
{v_T}_\alpha &=& \sqrt{\frac{2 k_B T_\alpha}{m_\alpha}}  \nonumber \\
{v_{\rm rel}}_{\alpha\beta} &=& \mid \vec{v_\alpha} -  \vec{v_\beta} \mid
\end{eqnarray}

\noindent
For the simulation presented in this paper we consider the viscosity of neutrals and charges and the thermal conductivity of the neutrals.
The viscosity is included in the pressure tensor:
\begin{equation} 
{\bf\hat{p}}_\alpha =  p_\alpha \mathbb{I} - {\bf\hat{\tau}}_\alpha,
\end{equation}
where the elements of  the viscosity tensor are:
\begin{equation} \label{eq:visc}
{\tau_\alpha}_{ij} =  \xi_\alpha \left ( \frac{\partial {v_\alpha}_i}{\partial x_j} + \frac{\partial {v_\alpha}_j}{\partial x_i} \right ) \,.
\end{equation}

\noindent
The viscosity and thermal conductivity coefficients are expressed as \citep[derived by][]{1965Braginskii}:

\begin{eqnarray} \label{eq:visc_th_coef}
\xi_\alpha &=& \frac{\sqrt{\pi k_B T_\alpha  m_\alpha}}{4 \Sigma_{\alpha\alpha}} \nonumber \\
K_\alpha &=& \sqrt{ \frac{\pi k_B T_\alpha}{m_\alpha}}\frac{1}{4 \Sigma_{\alpha\alpha}}\,,
\end{eqnarray}
and  depend only on the temperature.
\noindent The collisional cross sections used were:
$\Sigma_{nn} = 2.1 \times 10^{-18} \text{m}^{2}$,
$\Sigma_{in} = 1.16 \times 10^{-18} \text{m}^{2}$, and $\Sigma_{en} =   10^{-19} \text{m}^{2}$.
For the viscosity of the charges only ion-ion collisions are taken into account:
$ \Sigma_{ii} = \frac{40\pi}{3} \left ( \frac{e^2}{4\pi\epsilon_0 k_B T_c} \right )^2$,
where $e$ is the elementary charge, and $\epsilon_0$ the permitivity of free space.

The setup of the prominence thread simulation is described as L1-WN in Table~1 in \cite{rti1}. The numerical box of the size 2 $\times$ 8 Mm is covered by 2048 $\times$ 8192 grid points, which is 4 times higher resolution than in our previous work.
For completeness we repeat here the initial conditions:
\citep{rti1,Leake2014},
\begin{align} 
n_{\rm i0} &= n_0 \text{ exp} \Big( -\frac{z}{H_c} \Big), \label{eqs:rti_setup_ni} \\
n_{\rm n0} &= n_{\rm n00} \text{sech}^2 \Big(2 \frac{z}{L_0} - 1 \Big) + n_{\rm nb}, \label{eqs:rti_setup_nn} \\
T_0 &= T_b f(z), \label{eqs:rti_setup_T}  \\ 
B_0 &= B_{00}  \Bigg\{1+ \beta_p  \bigg[\frac{n_{\rm i0}}{n_0}  \Bigg(1-f(z)\Bigg) -  \frac{1}{2}  
\frac{n_{\rm n0}}{n_0}  f(z) \nonumber \\    
 & -  \frac{1}{H_c  n_0}   \Bigg(\frac{1}{2}  L_0  n_{\rm n00}  \text{tanh}
\Big(2 \frac{z}{L_0}-1\Big) + n_{\rm nb}  z \Bigg)  \bigg]\Bigg\}^{0.5},  \label{eqs:rti_setup_B} \\
p_{\rm n0} &= n_{\rm n0} k_{\rm B} T_0\,, \nonumber  \\
p_{\rm c0} &= 2 n_{\rm i0} k_{\rm B} T_0\,,\label{eqs:rti_setup_pe}
\end{align}
where the ion number density at $z=0$, $n_0$ = $10^{15}$ m$^{-3}$, the peak neutral number density,
reached at $z=L_0/2$, is $n_{\rm n00}$ = $10^{16}$ m$^{-3}$, 
the background temperature of the corona, $T_b$ = 2.02 $\times$ 10$^5$ K, the neutral number density corresponding to the corona temperature ($T_b$), $n_{\rm nb}$ = 3.5 $\times$ 10$^9$ m$^{-3}$, $B_{00}=10^{-3}$ T, the charges gravitational 
scale height, $H_{\rm c} = 2 {k_{\rm B} T_b}/{(m_{\rm H} g)}$. { The characteristic length scale is $L_0$=1~Mm.}
The plasma $\beta_p = {2 n_0 k_{\rm B} T_b}/({B_{00}^2}/2\mu_0)$ is calculated at $z=0$ using  $B_{00}$ as the value of the magnetic field, 
and has the value  $\beta_p \approx$  1.4 $\times$ 10$^{-2}$. 

{ The temperature profile, shown in Eq.~(\ref{eqs:rti_setup_T}), is described using the function $f(z)$:
\begin{equation}
f(z) = \frac{\text{cosh}^2\Big(\frac{z}{L_0}-0.5\Big)}{\Big(\frac{z}{L_0}-0.5\Big)^2 + L_{\rm t}}\,,\label{eq:f_z}
\end{equation}
with the value of $L_{\rm t}$ = 20 chosen so that the temperature is closest to the values observed in the Sun and the ionization fraction 
$\xi_{\rm i} = {\rho_{\rm c}}/{(\rho_{\rm c} + \rho_{\rm n})}$ = 0.091  remains small \citep[see][]{Leake2014}.}

The sheared magnetic field is contained initially in the $xy$ plane, with the variation only in the $z$-direction:
\begin{align} \label{rti_shear_setup}
&B_{\rm x0} = B_0 \text{sin} (\theta),\,\,\, B_{\rm y0} = B_0 \text{cos} (\theta) \text{, with:}\nonumber \\
&\theta(z) = \text{tanh}\left(\frac{z}{L_{\rm s}}\right) \theta_0 \times \pi/180 \,\,,
\end{align}
\noindent
where the values used in the simulation are 
$L_s$=1 Mm and $\theta_0=1^\circ$.

The boundary conditions are periodic in the $x$-direction.  In the $z$-direction, we use antisymmetric boundary conditions for the vertical velocity of charges and neutrals and symmetric for the rest of the variables.  We note that the boundaries in the z-direction are located sufficiently far from the prominence thread so as not to impact the dynamics of the simulation.
}

\section{Magnetized RTI dynamics and magnetic field amplification.}
\label{section3}
\begin{figure*}[!htb]
 \centering
 \includegraphics[width=18cm]{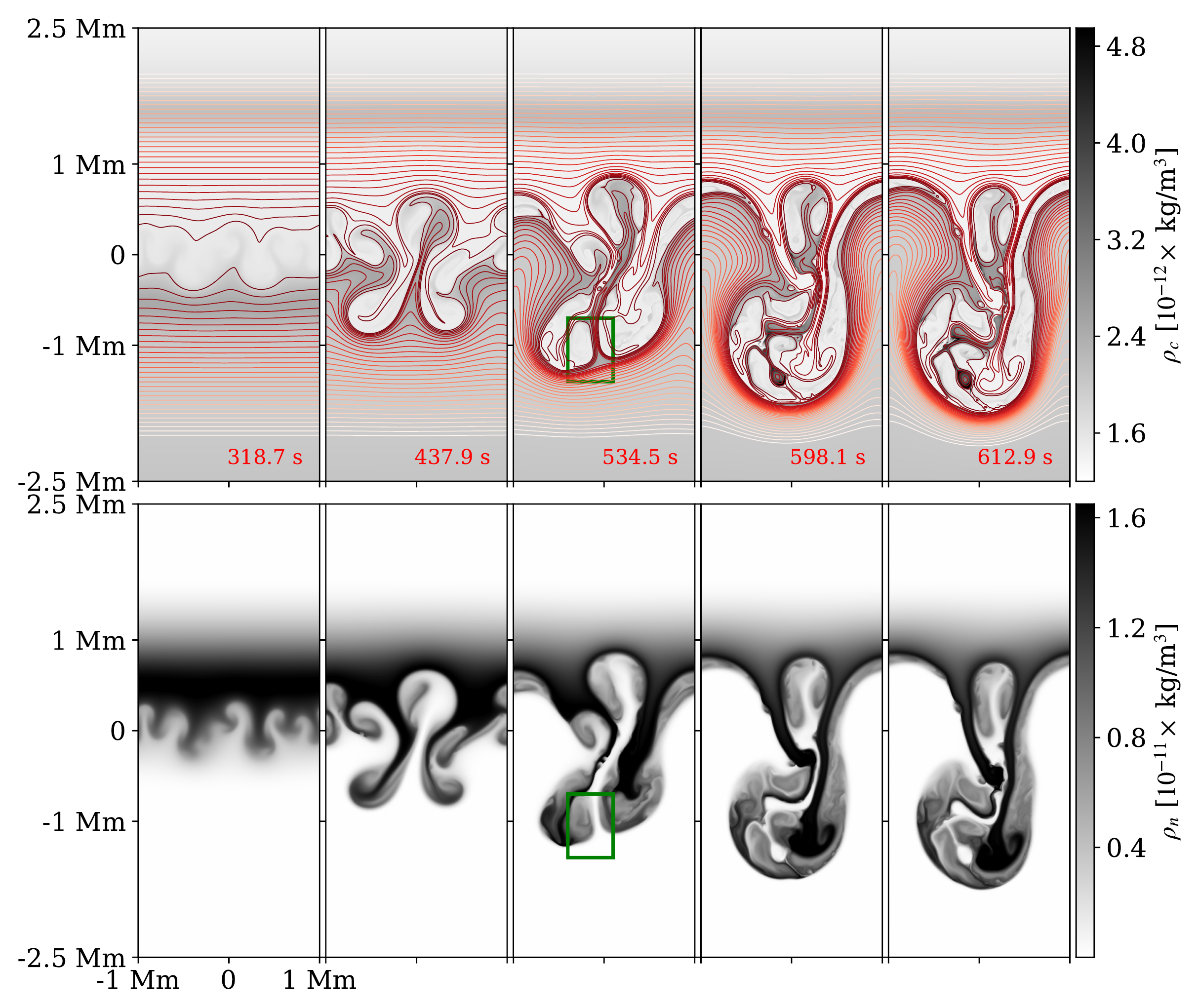}
\caption{Snapshots of the non-linear evolution of the RTI at $t=[318.7~s, 437.9~s, 534.5~s, 598.1~s, 612.9~s]$ for $(x,z)\in[-L_0,L_0]\times[-2.5~L_0, 2.5~L_0]$.  The top panels illustrate the evolution of the in-plane projection of the magnetic field lines overlaid on a grayscale colormap of the mass density of the charges.  The bottom panels similarly show the spatial distribution and dynamical evolution of the mass density of the neutrals.  The rectangular box marked in the $t=534.5~s$ panels corresponds to the zoom-in views of a reconnecting current sheet shown in Figs.~\ref{fig:CS_sequence} and \ref{fig:CS_image} below.}
\label{fig:time_snaps2_e}
\end{figure*}

\begin{figure*}[!h]
 \includegraphics[width=9.5cm]{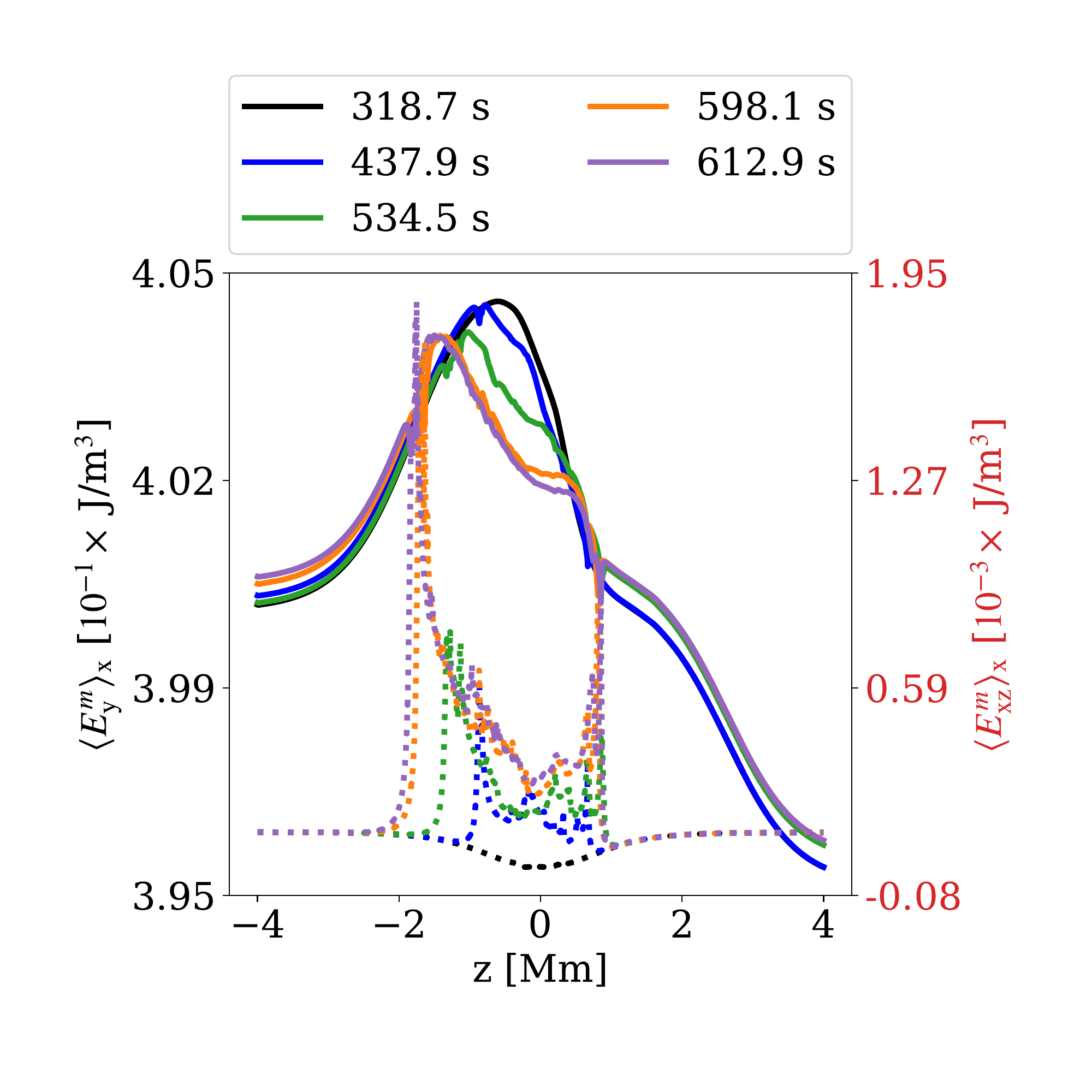}
 \includegraphics[width=8cm]{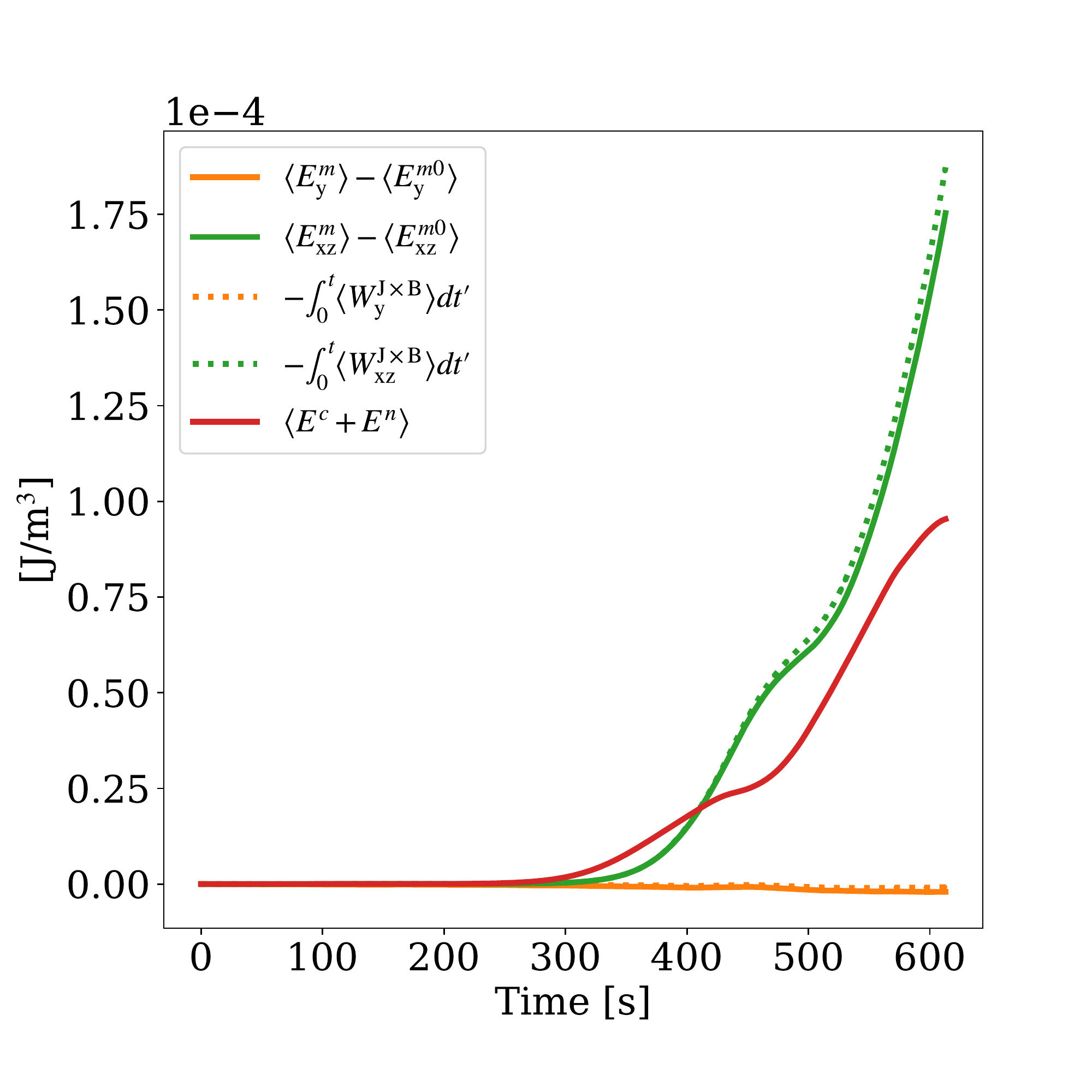}
 \caption{Left panel: Out-of-plane magnetic energy component (solid lines, values on the left axis) and the in-plane magnetic energy (dotted lines, values on the right axis), as a function of height at several time moments. The values are averaged in horizontal direction. Right panel: time evolution of the spatially averaged in-plane magnetic energy (green solid line), out-of-plane magnetic energy (orange solid line) and the sum of the kinetic energies of charges and neutrals (red line), taken with respect to $t=0$ values. Green dotted and orange dotted lines shows the work done against the Lorentz force by the in-plane and out-of-plane magnetic field components, correspondingly. }
\label{fig:Em_z_time}
\end{figure*}

We show in Figure~\ref{fig:time_snaps2_e} the time evolution of the Rayleigh-Taylor instability from the early non-linear through the fully developed phase.  Non-linear development of RTI leads to formation of downflows (spikes) and upflows (bubbles). The mass density of the charges (top row) and the mass density of the neutrals (bottom row) show similar isodensity contours, with regions of high neutral density corresponding to regions of low density of charges. 
As bubbles and spikes form, the in-plane magnetic field lines are dragged and stretched by the motion of the plasma. This is shown in the top panels of Figure~\ref{fig:time_snaps2_e},  where the magnetic field lines are plotted over the colormap of density of the charges. We observe the field lines to follow the isodensity contours, and observe a significant increase in the magnitude of the in-plane field, illustrated as the density of field lines, enveloping the spikes and upflows over time. Around time 534.5~sec (middle panels) one can observe that two bigger drops are brought together by the flow and start merging. This merging results in formation of one single spike, visible in the last two panels, which has complex internal structure. In particular, one can observe formation of current sheets and plasmoids. One of the bigger plasmoids is clearly visible at the last panel as a dark/bright circular structure in the charged(top)/neutral(bottom) density images. It is filled with the hotter and more ionized coronal-like material.

The large scale structures developed by the instability drag the in-plane magnetic field lines and we expect the in-plane magnetic energy to be concentrated around these structures. The generation of in-plane magnetic energy by the RTI is quantified in Figure~\ref{fig:Em_z_time}. The left hand side panel of Figure~\ref{fig:Em_z_time} shows the horizontal average of the out-of-plane magnetic energy, ${\langle E_{\rm y}^m\rangle}_{\rm x}$, (solid lines, numbers on the left axis) and in-plane magnetic energy, ${\langle E_{\rm xz}^m\rangle}_{\rm x}$, (dotted lines, numbers on the right axis) as a function of height for the same time moments as in Figure \ref{fig:time_snaps2_e}.  We observe  that the amount of in-plane magnetic energy rapidly grows as RTI develops, with concentrations of energy immediately above and below the bubbles and spikes. The two peaks in the magnetic energy, located at the height of spikes and bubbles, steepen.  The out-of-plane magnetic energy is redistributed within the prominence thread region, but does not appear to change significantly.  In this last phase the small scales dynamics dominates the evolution, similar to the results of \cite{Hillier2016} and to the time evolution shown in Figure~\ref{fig:time_snaps2_e}, where the height reached by the large scale structures barely changes in the last two snapshots.

The right hand side panel of Figure~\ref{fig:Em_z_time} shows the time change in the the spatially averaged out-of-plane magnetic energy, with respect to the initial one at $t=0$, computed as $\langle E_{\rm y}^m \rangle-\langle E_{\rm y}^{m0} \rangle$ (indicated by solid orange line). Similarly, it also shows the time change in the in-plane magnetic energy, $\langle E_{\rm xz}^m  \rangle - \langle E_{\rm xz}^{m0} \rangle$, indicated by solid green line. For comparison, the time evolution of the sum of the kinetic energy density of charges and neutrals, $E^c + E^n$ is shown in red solid line. We note that the initial free energy is mostly in the gravitational energy of the prominence thread, with high neutral density supported by a small increase in magnetic field strength at the prominence-corona interface.  The right hand panel of Figure~\ref{fig:Em_z_time} demonstrates that most of the free energy goes into the generation of the in-plane magnetic field.  There is a small decrease in the amount of out-of-plane magnetic energy, but it does not appear to be dynamically significant. While we observe an initial increase in the kinetic energy of neutrals, and to a smaller extent charges, as the bubbles and spikes accelerate, the in-plane magnetic field appears to contain and absorb plasma flows. At the conclusion of the simulation, we observe the increase in the magnetic energy (solid green line) to be approximately twice the kinetic energy of neutrals and charges (solid red line). 
{ In this last phase, as the small scales dynamics dominates the evolution, the magnetic energy continues to grow while the kinetic energy begins to saturate, consistent with the conclusion of \cite{Hillier2016}.

To verify our conclusions and estimate the amount of numerical dissipation impacting magnetic field generation, we also calculate the work done by RTI flows against both the in-plane, $-\int^{t}_0 \langle W_{\rm xz}^{J\times B} \rangle dt'$, and out-of-plane,  $-\int^{t}_0 \langle W_{\rm y}^{J\times B} \rangle dt'$, components of the Lorentz force. These two components of the work are computed as $W_{\rm xz}^{J\times B} = J_y \left( B_z {v_c}_x - B_x {v_c}_z\right)$ and $W_{\rm y}^{J\times B} = B_y \left( J_x {v_c}_z - J_z {v_c}_x\right)$, with 
$W^{J\times B} = \vec{F} \cdot \vec{v_c} = W_{\rm xz}^{J\times B} + W_{\rm y}^{J\times B}$. The results of these calculations are shown as greed dotted line (for $W_{\rm xz}$) and orange dotted line (for $W_{\rm y})$ at the right panel of Figure~\ref{fig:Em_z_time}. For both components, we observe that the corresponding magnetic energy curve shows slightly smaller values. This difference is due to the numerical magnetic diffusivity, as discussed below and quantified in Appendix~\ref{appen}. }


\section{Formation of current sheets, magnetic and plasma structures.}
\label{section4}
\begin{figure*}[!h]
 \centering
 \includegraphics[width=16.7cm]{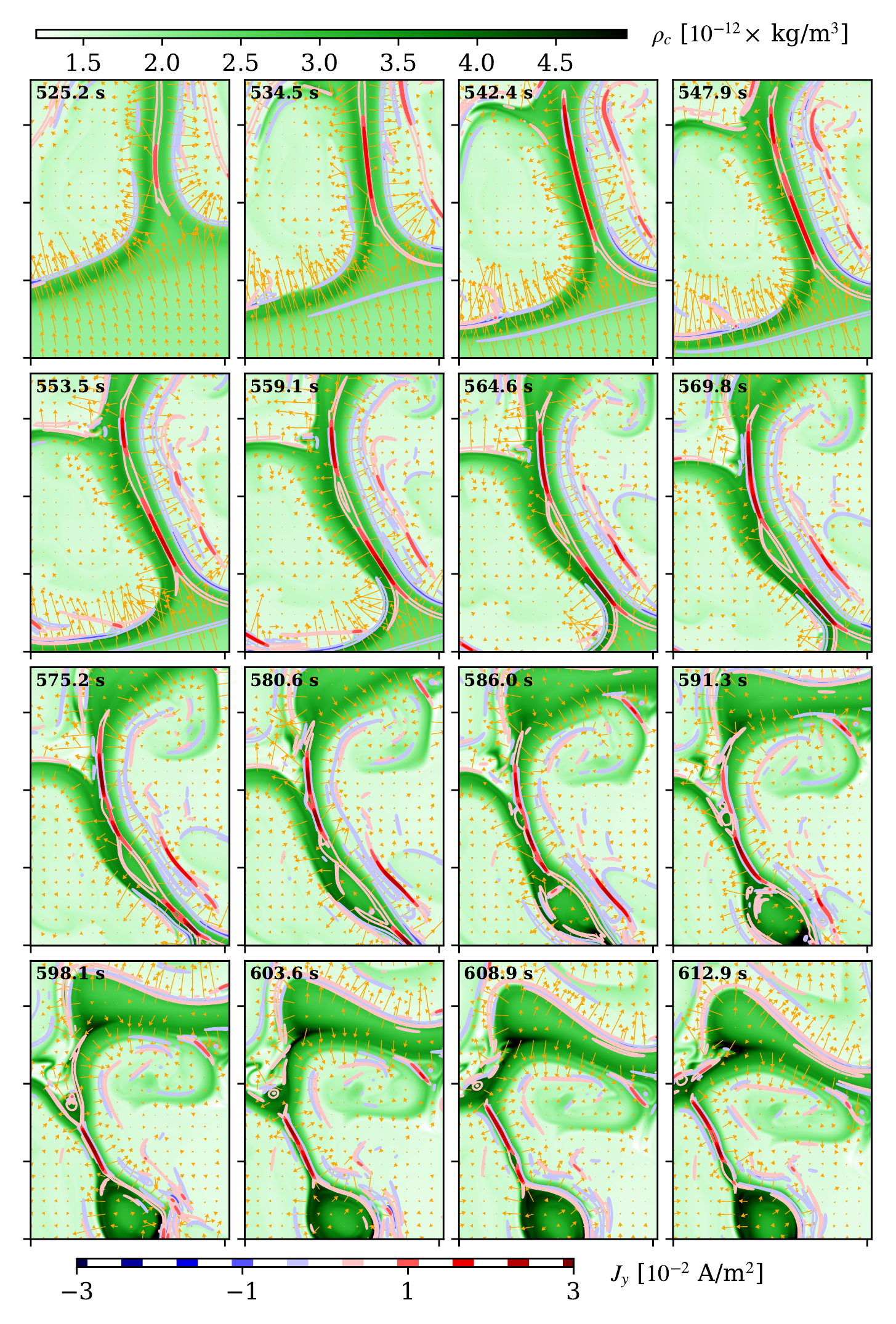}
 \caption{A snapshot sequence of dynamic evolution of a reconnecting current layer with recurring plasmoid formation is shown with color contours of the out-of-plane current density $J_y$ and arrows showing difference in the in-plane flow between charges and neutrals $(\vec{v}_c - \vec{v}_n)_{xz}$ plotted over a colormap of the mass density of charges. The snapshot window is illustrated in the $t=534.5~s$ panels of Fig.~\ref{fig:time_snaps2_e} and has coordinates $(x,z) \in [-0.4~{\rm Mm}, 0.1~{\rm Mm}] \times [-1.4~{\rm Mm}, -0.7~{\rm Mm}]$.  An arrow with length of 50~km corresponds to a speed difference of $|(\vec{v}_c - \vec{v}_n)_{xz}| = 128~\rm{m/s}$.}
\label{fig:CS_sequence}
\end{figure*}

\begin{figure*}[!ht]
 \centering
\includegraphics[width=16cm]{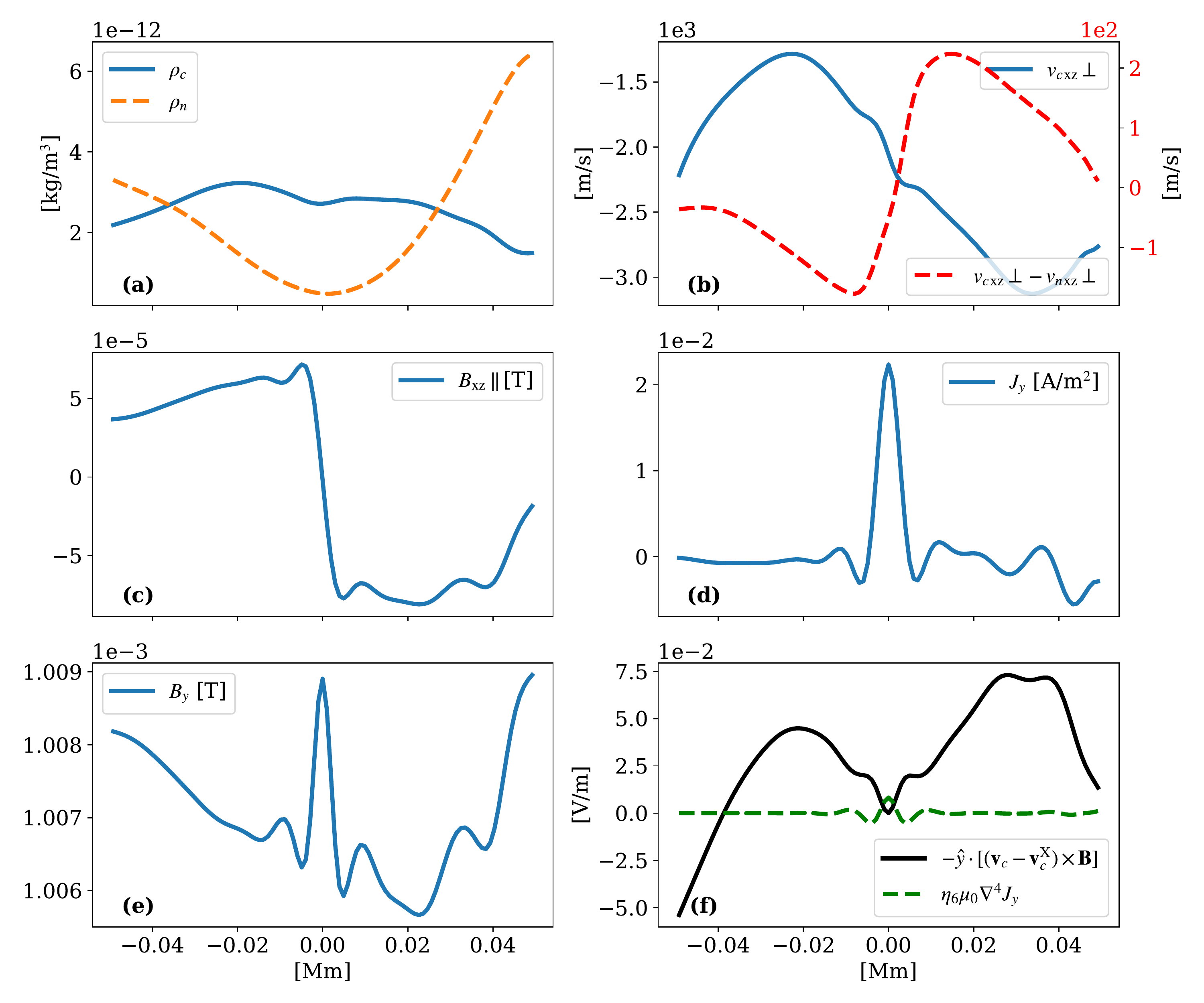}
\caption{Profiles across a reconnecting current sheet at $t=534.5~s$ along the cut marked with the dashed blue line, labeled ``$\perp$'' in left panel of Fig.~\ref{fig:CS_image} and centered with the $0$ location along the cut to correspond to the X-point location in Fig.~\ref{fig:CS_image}.  The panels show (a) mass density of neutrals (dashed orange line) and charges (solid blue line); (b) the flow velocity along the direction of the cut for the charges (solid blue line, left axis) as well as the relative drift between charges and neutrals (dashed red line, right axis); (c) the in-plane magnetic field component normal to the cut and along the current sheet; (d) the out-of-plane current density $J_y$; and (e) the out-of-plane magnetic field $B_y$. Panel (f) shows the calculated ideal electric field $E_y = -\hat{y}\cdot[(\vec{v}_c - \vec{v}^X_c)\times\vec{B}]$ in the frame of reference of the charged fluid at the X-point (solid black line), and the $\hat{y}$-component of the effective dissipative electric field, $\eta_6\mu_0\nabla^4 J_y$, present due to numerical dissipation (dashed green line). The width of the reconnection current sheet is measured to be $\lambda_J = 5.323~\rm{km}$ as the full width at half max of the $J_y$ profile in panel (d).}
\label{fig:CS_cut}
\end{figure*}

\begin{figure}[!ht]
 \centering
\includegraphics[width=8cm]{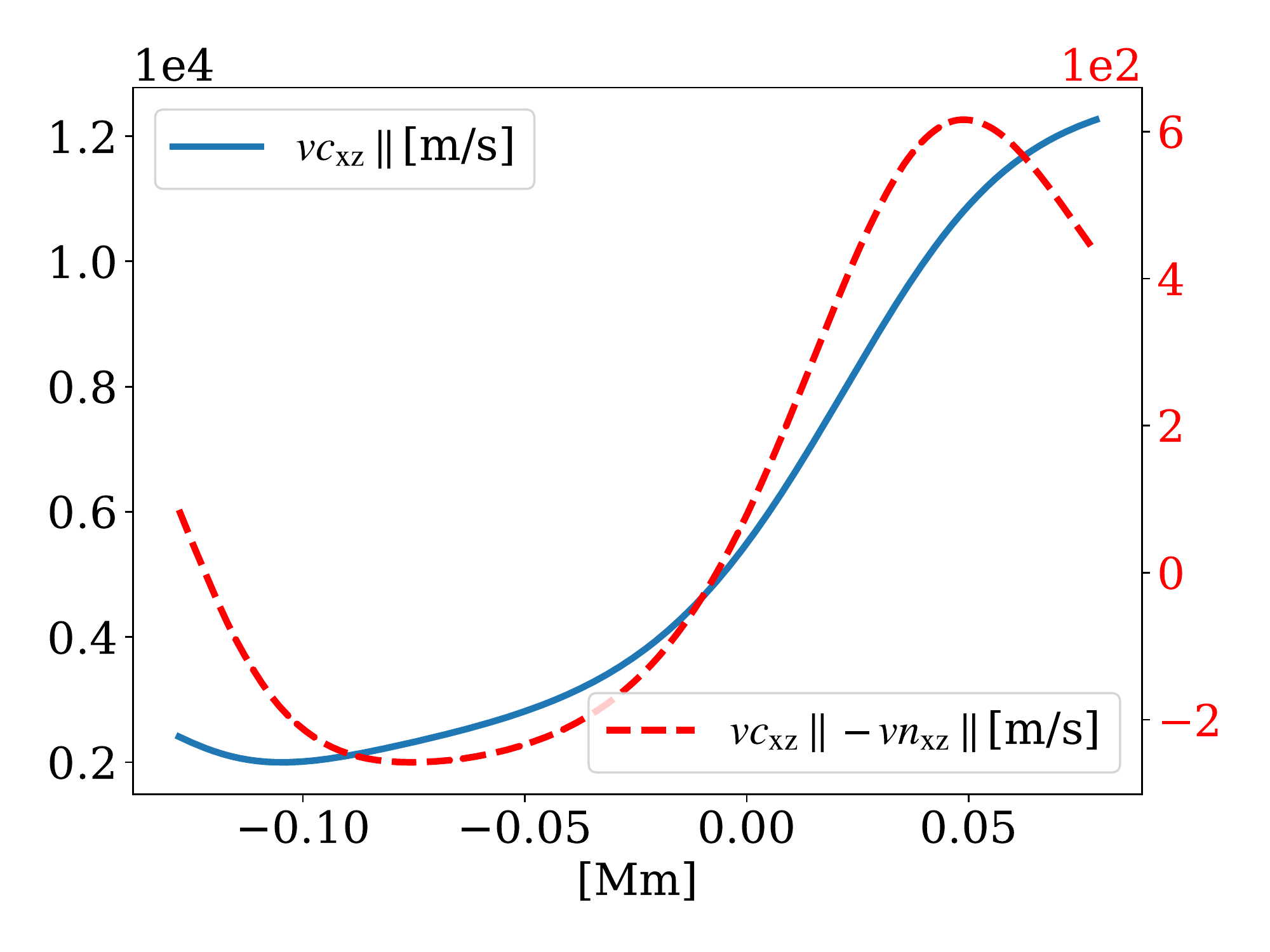}
\caption{ Outflow profiles along the reconnecting current sheet, the direction indicated by ``$\parallel$'' in Fig.~\ref{fig:CS_image} at time $t=534.5~s$. Blue solid line is for the charged velocity with values indicated at the left axis, red dashed line is for the decoupling velocity, with values indicated at the right axis. The location of 0 on the $x$-axis corresponds to the location the points marked by "X" along the current sheet in Fig.~\ref{fig:CS_image}. 
}
\label{fig:CS_cut2}
\end{figure}


\begin{figure*}[!htb]
 \includegraphics[width=8.5cm]{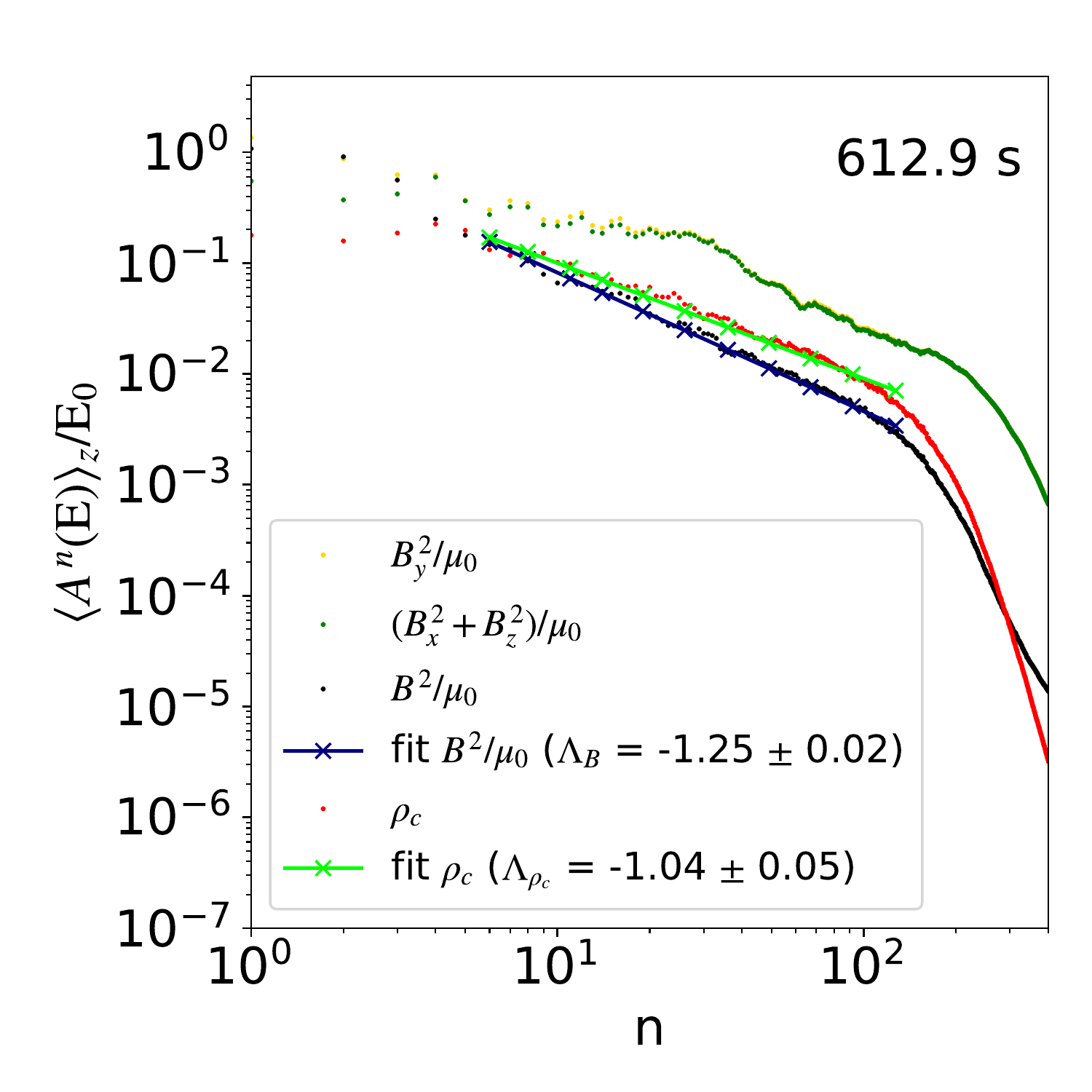}
 \includegraphics[width=8.5cm]{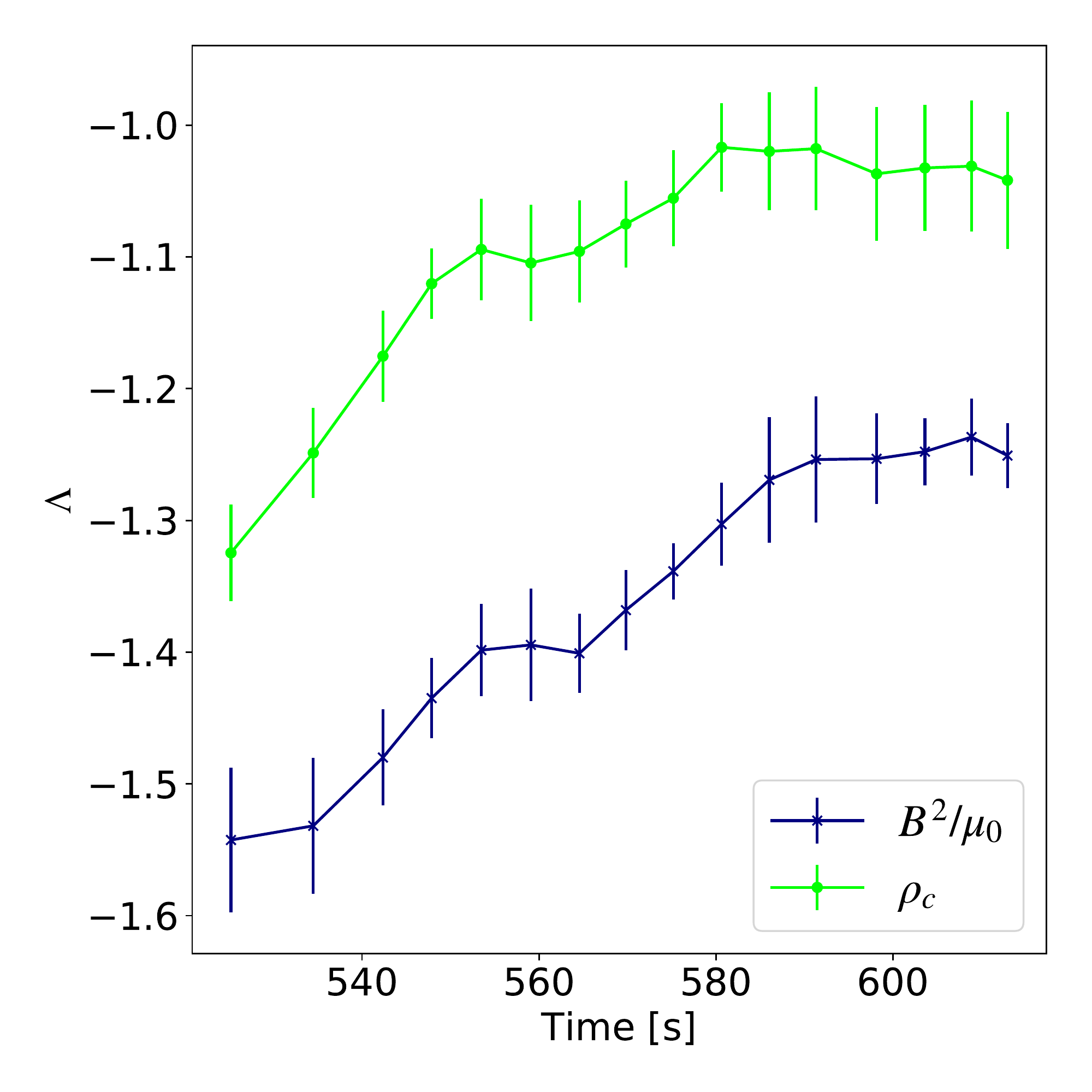}
 \caption{Left: Fourier decomposition of spatial structure in magnetic field and mass density of charges in the periodic $x$-direction at $t = 612.9~s$.  For each horizontal mode $n$, amplitudes are calculated and averaged in height over the region of high RTI activity between $z = -1.35~\rm{Mm}$ and $z = 0.65~\rm{Mm}$. Height-averaged Fourier decomposition amplitudes are shown for out-of-plane magnetic field, $B_y^2/\mu_0$, in-plane magnetic field, $(B_x^2+B_z^2)/\mu_0$, total magnetic field, $B^2/\mu_0$, and mass density of charges, $\rho_c$. The normalization value ($E_0$) used for magnetic field data is $10^{-3}$ J/m$^3$ and for $\rho_c$ is $10^{-12}$ kg/m$^3$.  The linear spectral fits $\langle A^n \rangle_z \propto n^\Lambda$ for $(B^2/\mu_0)$ and $\rho_c$ are calculated using a subset of 11 mode amplitude values for $n = [5, 7, 10, 13, 18, 25, 35, 48, 66, 91, 126]$ that are approximately equidistributed over the range of the fit in the $\log_{10} n$ space, as illustrated with 'x' markers in the panel.  Right: Time evolution of the spectral exponents $\Lambda_{B}$ and $\Lambda_{\rho_c}$ during the late non-linear and saturation phases of the RTI calculated at the same time instances as shown in the Fig.~\ref{fig:CS_sequence} snapshots. The errors of the fit are shown as vertical error bars.    
 }
\label{fig:magSp}
\end{figure*}

{ The magnetic field is frozen into the plasma fluid at large scales and the non-linear interaction between the spikes brings field lines of opposite polarity close to each other, thus creating current sheets. At small scales, magnetic diffusivity acts to reconnect the magnetic field lines.}

We recall that the simulations described here, and in \citet{rti1,rti2}, do not include an explicit resistivity term in the Ohm's Law and the magnetic diffusivity is numerical. The functional form of the numerical diffusivity and the value of the corresponding coefficient in the present \mancha implementation are described in Appendix~\ref{appen}.  In particular, we estimate that numerical magnetic diffusivity within observed current sheets for the simulation described in this paper is more than an order of magnitude greater than the physical diffusivity due to Spitzer resistivity would be.  Thus, given the practical limitations of a numerical simulation, we are justified in omitting the resistivity term from the Ohm's Law.

Figure~\ref{fig:CS_sequence} shows the evolution of charged fluid density in the window drawn in the middle panels of Figure~\ref{fig:time_snaps2_e}, which captures the two merging spikes. The current density contours are overplotted and we observe formation of current sheets associated with diffusion of the magnetic field.  As the system evolves, the current sheets repeatedly elongate and break up, leading to the formation of smaller secondary magnetic structures.  The series of snapshots in Fig.~\ref{fig:CS_sequence} documents at least five such instances, with the resulting small plasmoids themselves merging and several joining into a larger one seen at the bottom of the window.  The critical length-to-width ratio of the current sheets before they break up consistently appears to be $\approx 20$.  We also observe that the merging plasmoids lead to formation of not only magnetic but also density structures, with the electron density in the large resulting plasmoid significantly exceeding the ambient coronal electron density prior to RTI development.  

As the spikes filled with largely neutral material attempt to merge, the Lorentz force slows down the charged particles across the field lines while the neutrals are not affected by the magnetic field.  The decoupling in velocity between charges and neutrals is shown in Fig.~\ref{fig:CS_sequence} by orange arrows and appears largest around the edges of the spikes and across the current sheet.

In order to study a reconnecting current sheet in more detail, we choose a moment of time when a single current sheet is clearly defined.  We choose time t=534.5~s, which corresponds to the second snapshot in the sequence shown in Figure~\ref{fig:CS_sequence}.  We then project different quantities across the current sheet as shown in the 1D profiles in Figure~\ref{fig:CS_cut}. (The location and extent of the projection are shown in Figure~\ref{fig:CS_image} in Appendix~\ref{appen2}. See the dashed red line labeled ``$\perp$'' in  Fig.~\ref{fig:CS_image}.  The  projection was calculated to be in the direction perpendicular to the current sheet.)

The 1D profiles capture the outer inflow region of the reconnection current sheet, showing the two spikes consisting mainly of neutral material approaching each other.  This is reflected in panel~(a) of Fig.~\ref{fig:CS_cut}, with the neutral density shown with the dashed line being larger outside the current sheet, but diminishing within the current sheet with the plasma becoming close to fully ionized.  The edges of the spikes become ionized while moving into the much hotter corona and this fact is reflected in the two small peaks in the charged density outside the current sheet, shown with the solid line in panel~(a).  

Panel~(b) of Fig.~\ref{fig:CS_cut} shows that plasma is flowing into the current sheet from both left and right, while the overall motion of the spikes downwards and to the left leads to the drift of the whole current sheet system. { This inflow is driven by the neutral pressure gradients associated with the density profile shown in the panel~(a).} The decoupling in velocities, shown with the dashed line in the panel~(b), is significant and amounts more than 20\% of the inflow velocity. We note that the symmetric nature of the decoupling centered at the current sheet confirms that the decoupling is indeed associated with the reconnection process.  The neutrals, which are dominant in the spikes, are flowing into the current sheet faster, making the decoupling velocity between charges and neutrals point outwards in the current sheet.

The reversal of the reconnecting component of the magnetic field is illustrated in the panel (c) and the magnitude of the reconnection current, $J_y$, in the panel (d) of Fig.~\ref{fig:CS_cut}. 
{ The width of the reconnection current sheet, measured as the full width at half maximum of $J_y$ to be $\lambda_J = 5.323~\rm{km}$ 
is similar to the collisional mean free path between ions and neutrals, $\lambda_{\rm in}=2~\rm{km}$, calculated with the value of the plasma parameters at the reconnection point.  This affirms that the two-fluid effects are expected to be important.}
We note that the reconnecting field of $\approx 5\times 10^{-5}$~T is 5\% of the out-of-plane ``guide'' magnetic field shown in the panel (e).  We also note that the guide field peaks at the center of the reconnection current sheet.  An estimate of the variation in the reconnecting and guide field components of the magnetic field from the panels (c) and (e) shows that the magnitude of the magnetic field strength is approximately constant across the current sheet.  Thus, while the in-plane and out-of-plane magnetic field components both have spatial structure on the current sheet scales, the magnitude of B-field appears to have less structure on the smallest scales.

Panel (f) of Fig.~\ref{fig:CS_cut} shows the contributions to the out-of-plane component of the electric field in the frame of reference of the moving current sheet: the ideal contribution (solid black line) that drives the reconnection process, and the non-ideal contribution (dashed green line) due to the numerical diffusion, estimated in Appendix~\ref{appen}, that represents the resulting reconnection rate.  We note that the total $E_y$-field, i.e., the sum of the two contributions, has a minimum at the center of the current sheet, which is typical for driven magnetic reconnection \citep{driven-rec}.

Figure~\ref{fig:CS_cut2} shows the outflow velocity of charges, and the velocity decoupling along the current sheet at time t=534.5~s. We note that the outflow velocities are sufficiently large to be potentially detectable in observations. The magnitude of the outflow is comparable to the Alfv\'en speed, $v_A\approx 2 \times 10^4$~m/s, calculated using the value of the reconnecting field of $5 \times 10^{-5}$~T and the value of density of $4 \times 10^{-12}$~kg/m$^3$.  We observe that the decoupling in the outflow is approximately 2\% of the outflow velocity being much smaller than the decoupling in the inflow (by a factor of 10). This conclusion is similar to \cite{2012Leake,2013Leake} who find that the outflows are coupled, while there is significant decoupling in the inflow.

In the presented simulation, the reconnection current sheet plasmoids, as the current sheets themselves, form dynamically and self-consistently.  As shown in Figure~\ref{fig:CS_sequence}, the plasmoids break up the current sheets, which then reform and elongate until they reach the aspect ratio of $\approx$20 and are again disrupted by the next plasmoid formation.  This process re-occurring at multiple sites as RTI develops leads to structure formation in both magnetic field and plasma density.  The spatial correlation between the in-plane and out-of-plane magnetic structures can be observed in the spectra of component contributions to magnetic energy shown for the latest time of the simulation in the left panel of Figure~\ref{fig:magSp}. At intermediate and small scales, the content of magnetic energy is very similar for the in-plane and out-of-plane components, with much less content in the total magnetic energy.  This is consistent with the magnetic field profiles across the current sheet shown in Figure~\ref{fig:CS_cut} and the conclusion reached in \citep{rti1} for small scale magnetic structures.

 The left panel of Figure~\ref{fig:magSp} also shows the structure spectrum for mass density of the charged fluid.  A linear fit in the intermediate scale range is done for the spectra of the total magnetic energy (blue) and the charged fluid density (light green), showing that the two are distinct within the standard deviation of the fit procedure described in the figure caption. The right panel of Figure~\ref{fig:magSp} demonstrates that towards the end of the simulation the slopes of both spectra evolve to a stationary value, indicating that the simulation appears to have reached a fully developed non-linear RTI state. 
 
\section{Conclusions.}
\label{section5}


The results described above demonstrate that fully developed RTI in a weakly sheared magnetic field background can act to transfer the bulk of the released gravitational energy into amplification of magnetic field with associated magnetic and plasma structure formation on multiple scales.  The mixing between neutral and ionized fluids leads to stretching of magnetic field lines around RTI fingers that begin to interact via current sheets. The current sheets, in turn, become unstable to secondary instabilities that lead to further magnetic and plasma structure formation, including generation and merging of magnetic flux ropes, enhanced plasma heating, and formation of highly ionized plasma structures with electron density significantly above ambient corona electron density. { This multi-scale process is strongly influenced by two-fluid ion-neutral effects at the smallest scales associated with diffusion and magnetic reconnection.}

As described in \cite{rti1,rti2}, the linear and early non-linear evolution of magnetized RTI in a sheared magnetic field can strongly depend on the shear scale relative to the density gradient scale driving the RTI.  In this paper, we focused on a prominence thread configuration with the same magnetic shear and density gradient scales.  This configuration allows simultaneous growth and development of several RT modes, which leads to robust mixing between the high density and dominantly neutral prominence material and the low density, magnetized, and well-ionized coronal plasma.  At the same time, due to the stabilizing presence of the sheared magnetic field, there is limited plasma mixing on short spatial scales, and the potential gravitational energy initially stored in the high density neutral material is largely deposited into the magnetic field energy by stretching the magnetic field lines on intermediate spatial scales.
{ 
We note that the conclusions presented here are limited by the fact that we use a 2.5D geometry.  The growth of and interaction between different modes in a fully three-dimensional system are likely to introduce additional complexity to the non-linear evolution of RTI \citep{Stone2007, Hillier2016}.
}

The interaction between magnetic field structures resulting from the early non-linear RTI development inevitably leads to formation of current sheets and associated dissipation of the magnetic energy.  While the set of two-fluid MHD equations evolved in this study does not include explicit dissipation terms in the Ohm's Law, we demonstrate that the 6th order spatial filtering procedure applied during the temporal advance of the PDEs in the \mancha code is the primary source of the dissipation of the magnetic field in current sheets.  

To estimate the effective diffusion due to two-fluid effects, we calculate the value of the ambipolar coefficient as:
\begin{equation}
\eta_A = \frac{\rho_n B^2}{\rho_c (\rho_c+\rho_n)^2 \alpha }.
\end{equation}
We note that while ambipolar diffusion does not constitute a formal dissipation mechanism for the magnetic field, comparing the magnitude of ambipolar diffusion to that for magnetic field dissipation due to numerical filtering provides a good measure of the relative impact of the two.  Using the value of the magnetic field of $B=5 \times 10^{-5}$~T, and the values for the collisional parameter $\alpha$ and the density of charges and neutrals at the center of the current sheet studied in Figure~\ref{fig:CS_cut}, we get a value $\eta_A = 2.19$~$\Omega$~m. This is three orders of magnitude larger than the value of the magnetic numerical dissipation, as calculated in Appendix~\ref{appen}., and reaffirms the necessity of two-fluid modeling of the system under consideration.

Magnetic reconnection between RTI-generated magnetic structures observed in the simulation is driven by the convective flows of the RTI fingers and the neutral fluid pressure gradient at their edges.  It is driven two-fluid magnetic reconnection with strong guide field, with neutral flows pushing the magnetic fields with oppositely directed in-plane components towards each other and producing flow decoupling between the neutral and charged inflows at $\approx 20\%$ of the neutral inflow speed. { This neutral flow driven two-fluid reconnection regime is distinctly different from previously considered cases of magnetically driven reconnection in two-fluid partially ionized plasmas \citep{2012Leake,2013Leake,Murtas_2021}.} Concurrent ionization of the inflowing neutral prominence material due to higher ambient temperature of the plasma between RTI fingers, as well as dissipative heating, results in current sheets being embedded within plasma structures with high ionization fraction and electron density.

Unlike prior work on two-fluid guide field reconnection in a weakly ionized plasma by \cite{Ni2018,Ni2018a}, the ionization and recombination processes do not play a dynamical role in determining the sub-structure of the reconnection current sheet in the present work.  This is due to several factors, including the ambient plasma parameters with lower ionization and recombination rates relative to the current sheet formation time, and the driven rather than spontaneous nature of the reconnection process.  It is also important to note that the present model does not include optically thin radiative losses included in the two-fluid model used by \cite{Ni2018} and \cite{Ni2018a}, which was shown in \cite{Ni2018a} to strongly impact the temperature and density current sheet sub-structure in some plasma regimes.

The driven reconnection process results in recurring formation of small scale magnetic flux ropes within the current sheet, often referred to as plasmoids, when the current sheet length-to-width aspect ratio approaches $\approx 20$, a value that is a factor of several lower than that reported by \cite{Ni2018b}.  While the numerical nature of the dissipation in the present simulation makes it difficult to generalize this result to other driven reconnection systems, this appears to be a robust feature of the present system as demonstrated in Fig.~\ref{fig:CS_sequence}. 
Further dedicated effort to explore stability of driven current sheets in a partially ionized plasma would be necessary to make more definitive conclusions regarding specific current sheet aspect ratio stability thresholds.

The RTI-driven plasma and magnetic field mixing, formation of current sheets within well-ionized plasma layers between RTI fingers, and the recurring formation of secondary magnetic and plasma substructures naturally leads to the development of a spectrum of structures at different scales.  We demonstrate that fully developed magnetized RTI can lead to emergence of { stationary} power spectra for structures in magnetic pressure and density of the charges (or electron density) in what may be analogous to the inertial range in turbulence.  We further show that the power spectra for the two quantities are different, though interpretation of the specific power spectra obtained is beyond the scope of this paper and is subject for future work. 

High resolution observations of prominences have been used to construct the power spectra of the structures over the regions with RTI dynamics \citep{2017A&A...597A.111H, 2016ApJ...818...57F, 2018ApJ...866...29F}, with the attention focused mostly on the velocity data. Assuming observed intensity to be a proxy for the plasma density, using different spectral lines for observations, e.g. neutral hydrogen H$_{\alpha}$ line or a line of an ionized element, e.g. CaIIH, one could also explore differences in density structure power spectra for plasma elements at different temperatures. As for the magnetic field power spectra, these require much more demanding spectropolarimetric observations and inversions. However, even maps of circular or linear polarization spectra can be used as a proxy. Next generation of the 4-m class solar telescopes (DKIST, EST) should allow for such analysis in the future. It will allow for better understanding of the relationship between the electron density structure spectrum and the magnetic pressure structure spectrum, and may enable identification of the range and spectrum of magnetic structure scales present at solar prominence - corona interfaces.

\begin{acknowledgements}
This work was supported by the Spanish Ministry of Science through the project PGC2018-095832-B-I00 and the US National Science Foundation. The work was also  supported by the FWO grant 1232122N.  It contributes to the deliverable identified in FP7 European Research Council grant agreement ERC-2017-CoG771310-PI2FA for the project ``Partial Ionization: Two-fluid Approach''. The author(s) wish to acknowledge the contribution of Teide High-Performance Computing facilities to the results of this research. TeideHPC facilities are provided by the Instituto Tecnol\'ogico y de Energ\'ias Renovables (ITER, SA). URL: http://teidehpc.iter.es . Any opinion, findings, and conclusions or recommendations expressed in this material are those of the authors and do not necessarily reflect the views of the US National Science Foundation.
\end{acknowledgements}

\bibliographystyle{aa}

\clearpage
\appendix

\section{Evaluation of numerical dissipation}
\label{appen}
\begin{figure*}[]
 \centering
\includegraphics[width=8.3cm]{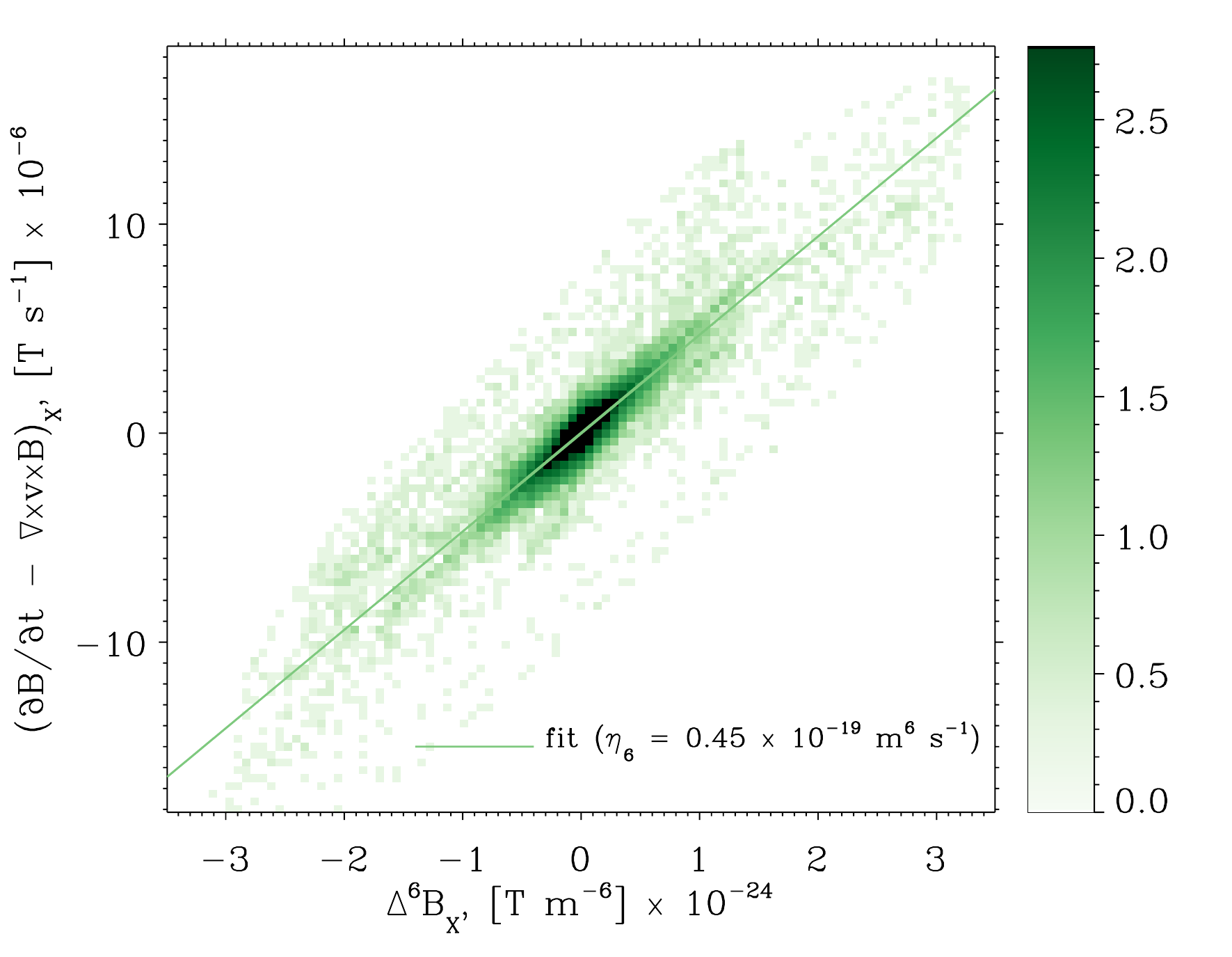}
\includegraphics[width=7.7cm]{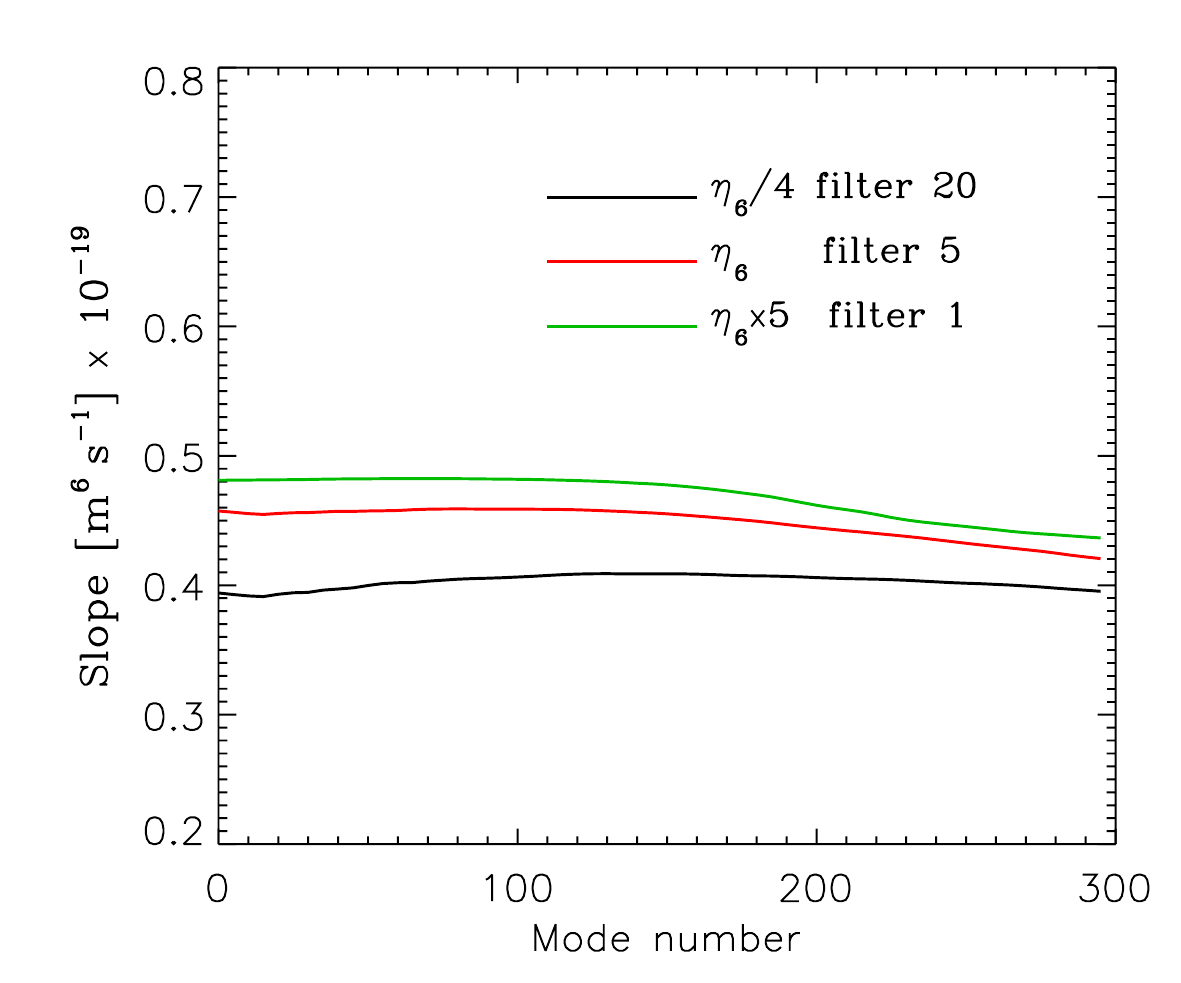}
\caption{Left: Scatter plot showing the dependence between the terms of the induction equation, $y={\partial {\mathbf B}}/{\partial t} - \nabla\times v\times{\mathbf B}$ (vertical axis) and $s=\Delta^6{\mathbf B}$ (horizontal axis). The intensity of the color indicated the number of spatial data points with the given values of (s, y) in log$_{10}$ units. The green line shows the linear fit to the distribution, with the value of the slope $\eta_6^F$ indicated in the figure. Right: dependence of the slope $\eta_6^F$ on the spatial mode number, obtained after the fitting to the simulation data (as those from the left panel) for three simulations with the different values of the filtering frequency: every 1 time step (green line), every 5 time steps (red line), and every 20 time steps (black line). The dependence on the mode number is obtained after applying a 2D Fourier high-pass filter to $y={\partial {\mathbf B}}/{\partial t} - \nabla\times \left(v\times{\mathbf B}\right)$ and $s=\Delta^6{\mathbf B}$ variables in each case. }
\label{fig:etafit}
\end{figure*}


The simulation studied in this work was performed with idealized Ohm's law and strictly zero Ohmic diffusion coefficient. Likewise, the artificial analogues of the numerical magnetic diffusivity, viscosity and conductivity were all set to zero. The only stabilizing method employed was the filtering.  Therefore, the filtering provides the numerical dissipation mechanism that allows for dissipation of the currents and magnetic reconnection in our model. In order to evaluate the amount of this numerical dissipation we consider the numerical details of how the filtering acts. The filter employed in \mancha is a sixth-order digital filter following  \cite{2007ApJ...666..547P}. It is based on the filtering function $G(k\Delta x)=1-\sin^6\Big (\frac{k\Delta x}{2}\Big )$. The frequency of application of the filter in this simulation was every 5 time steps.

If one considers that the application of the filter at a given time step consists in changing a variable by
\begin{equation}
\frac{\partial u}{\partial t}=\frac{(u(x)-u_{\rm filt}(x))}{\Delta t}=\sum_{m=-3}^3d_{\rm m} u(x+m\Delta x)/\Delta t,
\end{equation}
\noindent where $u$ is a variable before filtering and $B_{\rm filt}$ is after filtering, and $\Delta t$ is the time interval between two successive application of the filtering. The coefficients $d_{\rm m}$ take the values,
\begin{eqnarray}
d_m &=& [d_{-3}, d_{-2}, d_{-1}, d_{0}, d_{1}, d_{2}, d_{3}] \\ \nonumber
&=& [-1, 6, -15, 20, -15, 6, -1]/64.
\end{eqnarray}

\noindent Since the filter is 6th order, it introduces a 6th order dissipation, which can be related to the time-change of $u$ as,
\begin{equation}
\frac{\partial u}{\partial t}=\eta^F_6\Delta^6 u,
\end{equation}    
where $\eta^F_6$ is the dissipation coefficient we are looking for. Assuming a discrete 2nd order 6th derivative of $u$ on a symmetric stencil along the direction $x$,
\begin{equation}
\Delta^6 u=\frac{\partial^6 u}{\partial x^6}=\sum_{m=-3}^3c_{\rm m} u(x+m\Delta x)/\Delta x^6,
\end{equation}
with $c_m=64d_m$, the coefficient $\eta^F_6$ can now be evaluated as:
\begin{equation}
\label{eq:eta6}
\eta^F_6=\frac{\Delta x^6}{64\Delta t}.
\end{equation}
In this estimate, $\Delta x$ is the size of our numerical grid. After introducing the values of $\Delta x$ and $\Delta t$ from simulations, the coefficient of the numerical  dissipation  obtained using Eq. (\ref{eq:eta6}) has the value $\eta^F_6 = 4.714 \times 10^{18}$ m$^6$ s$^{-1}$.

In order to check the correctness of the above order of magnitude estimate and to get a better understanding of the numerical dissipation caused by the filtering we have computed $\eta^F_6$ from the simulation data. For that, we have re-run the simulations for a short interval of time between 534.5 and 535.0 sec, applying the filtering either every 1, 5 or 20 time steps. Using these data, we made the linear regression between the corresponding terms in the induction equation, assuming 6th order dissipation,
\begin{equation}
    y=\eta^F_6 s +c,
\end{equation}
where
\begin{eqnarray}
    s&=&\Delta^6 {\mathbf B}; \\ \nonumber
    y&=&\frac{\partial {\mathbf B}}{\partial t} - \nabla\times \left(v\times{\mathbf B}\right).
\end{eqnarray}
The regression was made separately for $x$ and $z$ components of ${\mathbf B}$, giving similar results for the $\eta^F_6$. The results of this calculation are shown in Figure \ref{fig:etafit}. The left panel of this figure demonstrates a well-defined linear dependence between $y={\partial {\mathbf B}}/{\partial t} - \nabla\times \left(v\times{\mathbf B}\right)$ and $s=\Delta^6{\mathbf B}$, with a very narrow scatter of the data points, confirming that our numerical dissipation is well represented by the 6th order. The value of the dissipation coefficient, obtained after the linear fit to the data points, $\eta^F_6=4.5 \times 10^{18}$ m$^6$ s$^{-1}$, coincides extremely well with the order of magnitude estimate given by the Eq. (\ref{eq:eta6}). 

The right hand side panel of Figure \ref{fig:etafit} shows how the value of $\eta^F_6$, obtained after fitting the numerical data, depends on the frequency of application of the filter (lines of different color) and on the mode number of the structures in the simulations. In order to obtain the dependence on the mode number, we applied a 2D high-pass Fourier filter to $y(x,z)={\partial {\mathbf B}}/{\partial t} - \nabla\times \left(v\times{\mathbf B}\right)$ and $s(x,z)=\Delta^6{\mathbf B}$ variables calculated for a given snapshot, prior to computing the linear regression. We have retained all the modes with a mode number above a given value $n$, being $n$ the mode number shown at the horizontal axis of the right hand side panel of Figure \ref{fig:etafit}. One can appreciate that the dependence of $\eta^F_6$ on the mode number is rather weak, meaning that our numerical dissipation affects equally the scales up to $n=300$, i.e. down to $L=1/n \approx 6.8$ grid points. The values of $\eta^F_6$ also scale well with the frequency of application of the filter. After we have re-scaled the curves for the different filtering frequency, the resulting values vary between $4\times 10^{18}$ and $5\times 10^{18}$ m$^6$ s$^{-1}$, which is, again, very close to our order of magnitude estimate from  Eq.~(\ref{eq:eta6}). 

The effects of numerical diffusivity can be compared to the magnetic diffusivity due to physical Spitzer resistivity at a given spatial scale. For this purpose, we calculate the Spitzer resistivity as $\eta_\parallel = \eta_\perp$/1.96, with $\eta_\perp$ given in \cite{NRL}:
\begin{equation}
    \eta_\perp = 1.03 \times 10^{-4} Z \text{ln}(\Lambda) T_{\rm eV}^{-3/2}\, \Omega \, {\rm m}.
\end{equation}
Here $T_{\rm eV}$ is the temperature in eV units, $Z=1$ in our case, and  we consider $\text{ln} (\Lambda)$ = 10.
Using the temperature at the center of the current sheet $T=2.2 \times 10^4$ K, we obtain the value of $\eta_\parallel = 2.01
\times 10^{-4}$ $\Omega$ m.  Comparing the 6th order numerical diffusivity with the physical resistivity on the current sheet scale,  we consider the length unit $L$ as being the width of the current sheet measured from the simulation, i.e. $L=\lambda_J = 5.3$~km (see Fig.~\ref{fig:CS_cut}), and calculate $\eta_2 = \eta^F_6 \mu_0/L^4 = 7.51 \times 10^{-3}\Omega$~m~$\gg \eta_\parallel$.  We thus observe that the numerical dissipation on the current sheet scale is more than an order of magnitude greater than the expected physical dissipation, justifying the fact that the latter has been neglected in the simulation.

{ As an additional check on the numerical accuracy of the presented results, we measure the spatial average of the absolute value of the divergence of the (in-plane) magnetic field throughout the simulation.  We observe that the maximum value of this quantity normalized to $B_{\rm ref}/L$ is below $4 \times 10^{-4}$, where we consider $B_{\rm ref}=5\times10^{-5}$~T and $L=\lambda_J=5.323$~km to be the magnitude of the reconnecting field and the width of the current sheet, respectively.  We thus consider $\nabla\cdot\vec B$ to be sufficiently small to not impact any of the conclusions presented in this paper.}

\section{Image with the projection lines used for plotting quantities across and along the current sheet}
\label{appen2}
\begin{figure}[!t]
 \centering
 \includegraphics[width=8cm]{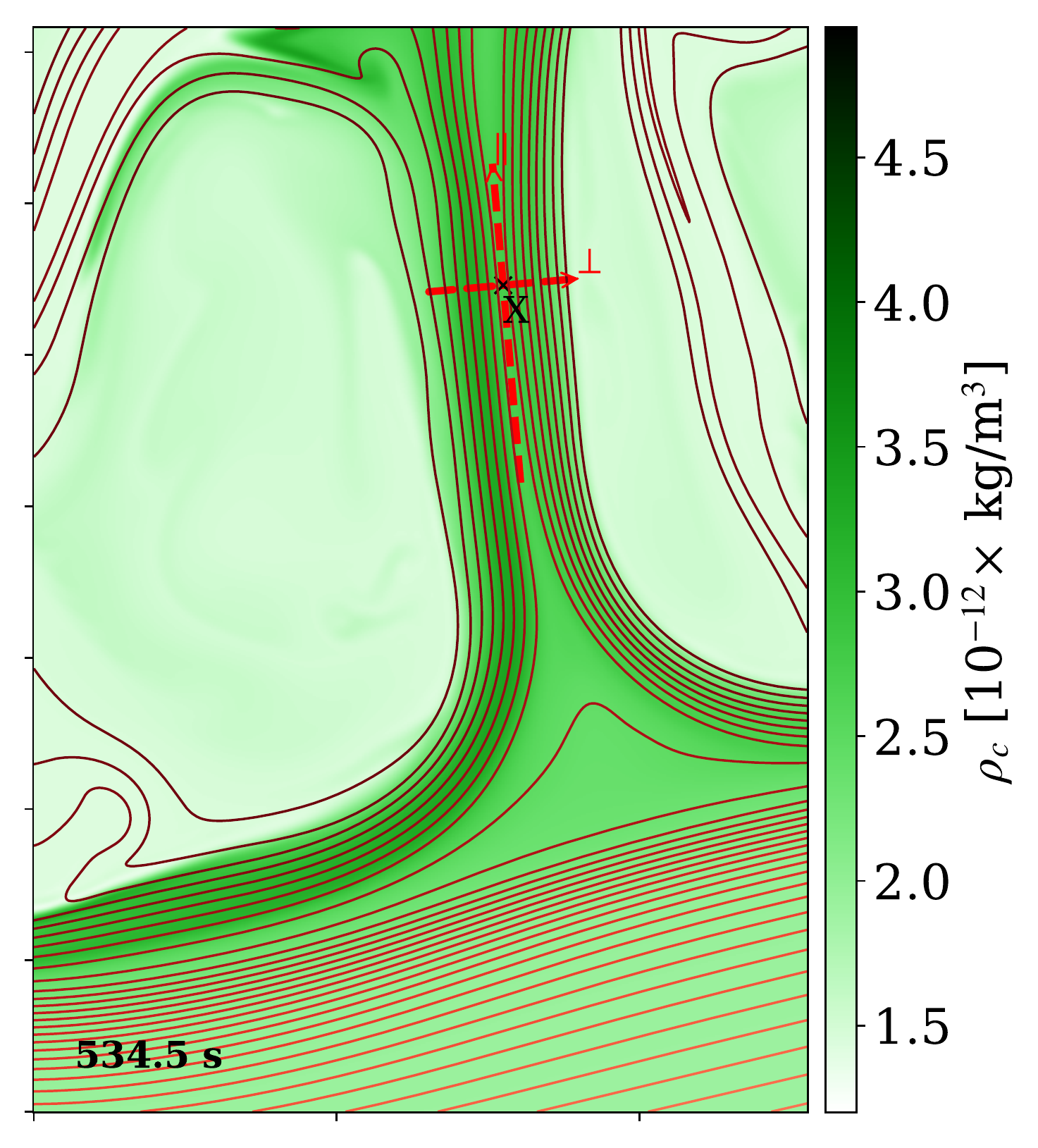}
 \caption{Snapshot of a reconnecting current sheet at $t=534.5~s$  with contours denoting in-plane projection of magnetic field lines, and a colormap of the density of the mass density of charges $\rho_c$. 
 The snapshot window is the same as in Fig.~\ref{fig:CS_sequence}.  
 The center x-point of the current sheet determined as the location of the maximum out-of-plane current density $J_y$ is marked with 'X' in the image.
 The dashed red line marks a cut along and across the current sheet in the direction parallel and perpendicular to the reconnecting magnetic field lines, respectively, as marked in the figure. }
\label{fig:CS_image}
\end{figure}

\end{document}